\documentclass{llncs}
\usepackage[T1]{fontenc}
\usepackage[utf8]{inputenc}
\usepackage{booktabs}
\usepackage{tabularx}
\usepackage{graphicx}
\begin{document}
\title{Prototyping Compliance: Participatory Legal UX for Platform Reporting Mechanisms under the DSA}
\titlerunning{Prototyping Compliance}
%

%
%
\author{Marie-Therese Sekwenz\inst{1} \and Daria Simons\inst{2} \and Alina Wundsam\inst{3}}
\institute{
  Delft University of Technology, Faculty of Technology, Policy and Management, Delft, The Netherlands\\
  \email{m.t.sekwenz@tudelft.nl}
  \and
  University of Amsterdam, Institute for Information Law (IViR), Amsterdam, The Netherlands\\
  \email{d.p.simons@uva.nl}
  \and
  IBM, Amsterdam, The Netherlands\\
  \email{alina.wundsam@ibm.com}
}

\maketitle              
\begin{abstract}

Digital regulations such as the European Union's Digital Services Act (DSA) represent major efforts to shape human-centered and human rights-based frameworks for society. Yet, as these laws are translated into practice, challenges emerge at the intersection of technology, law, and design. This paper presents a qualitative case study examining how designers act as mediators between abstract legal requirements and real-world digital experiences for users, focusing on the design of content reporting mechanisms under Article 16 of the DSA.

Through an expert workshop with professional designers from diverse fields (N=9), we explore how legal obligations are interpreted by designers and reflected in  discussions and design solutions. Our findings resonate with previous research on the design of reporting mechanisms and dark patterns, highlighting how UX design choices can mislead or hinder users' decision-making and therefore also highlighting the crucial role of design decisions.
 We show how participatory design methods can bridge disciplinary divides, making legal obligations accessible in compliance fostering design solutions.

By using legal design as a lens, we argue that the co-creation of digital regulations and user experience is a core site for digital humanism; where designers, engineers, and legal scholars must collaborate to ensure that systems uphold legal standards to address the challenge the regulation poses to these disciplines.

\end{abstract}
\section{Introduction}
The European Union's “Digital Decade” outlines an ambitious road-map toward a digitally sovereign, inclusive, and sustainable society \cite{noauthor_europes_nodate}. It is supported by a suite of legislative initiatives, including the Artificial Intelligence Act \cite{noauthor_regulation_2024}, the Data Act \cite{noauthor_proposal_2022}, the NIS2 Directive \cite{noauthor_directive_2022}, or the Digital Services Act (DSA) \cite{noauthor_regulation_2022}. Together, these regulations aim to address systemic challenges in the digital sphere; ranging from content moderation to algorithmic accountability, human rights-based audits, to digital governance. While these laws are grounded in the protection of fundamental rights \cite{charter_fundamental_rights_eu}, their real-world impact depends heavily on the ability of engineers and designers to interpret and implement them in systems.

For designers, this presents both a challenge and an opportunity. Unlike legal professionals, designers are rarely trained in legal interpretation. Yet they are increasingly responsible for embedding legal norms such as "privacy by design" \cite{tunca2024privacy,kostova2020privacy}, and legally demanded positioning in qualities like "user friendliness" into digital interfaces \cite{habib2020scavenger,sarkar2023should}. The design of user reporting systems enables users to speak up in content moderation systems of online platforms as a crucial access point for platform accountability \cite{jhaver2019human,mclean2019female}. The design of such reporting systems is now regulated under Article 16 DSA.

In this paper, we explore how designers engage with regulation through the case of content reporting system design. We present findings from an expert workshop with \emph{anonymized for review} designers, where we examined current reporting mechanisms on major platforms and asked participants to redesign them based on legal and usability heuristics. Therefore our research questions are: (RQ1) \emph{How do designers interpret regulatory goals in design practice?} (RQ2) \emph{What usability barriers exist in current systems, and how can design better support legal compliance?},and (RQ3) \emph{What lessons can be drawn from participatory legal UX work to inform design solutions in the light of digital humanism principles?}

This paper makes four contributions in response to our research questions. First, it provides empirical insights into how designers engage with abstract regulatory goals such as those embedded in the DSA (RQ1), showing that while legal familiarity is limited, designers intuitively draw on user-centered values like clarity, fairness, and accessibility when reviewing interfaces. 

Second, the paper identifies critical usability barriers in existing reporting systems (RQ2), including poor discoverability, legalistic language, lack of feedback, and undifferentiated user pathways. 

Third, it reflects on the role of participatory design methods such as persona-driven scenarios and heuristic evaluation as entry points for initiating legal sensitivity and interdisciplinary dialogue in design contexts in line with digital humanism principles (RQ3) \cite{dighum_manifesto2019}.

Fourth, our findings point toward a more prominent role for designers in shaping policy and law, not only by surfacing value tensions through their work but also by actively resolving them. 

Designers are uniquely positioned to translate regulatory intentions and abstract values into concrete, effective digital experiences (RQ 2-3). A function increasingly recognized in emerging fields such as Design for Values \cite{vandenhoven2015,spiekermann2015ethical,spiekermann2020valuebased}
. We propose that future research and policy should more explicitly engage designers as key actors in the development, translation, and testing of digital regulation, especially as AI-human collaborations and platform governance evolve.

\section{Background}
Recent years have witnessed growing interest in the intersection of law and design, with emerging fields such as legal design \cite{perrykessaris2019legaldesign}, legal UX \cite{dickhaut2022designpatterns}, legal informatics, and design for regulation gaining traction \cite{urquhart_legal_2022}. Scholars from different disciplines have emphasized the role of design in giving shape to normative concepts like justice \cite{10.1093/0195180992.003.0007}, transparency \cite{flyverbom2015transparency,flyverbom2019digital}, or accountability \cite{jhaver_does_2019}. Similarly, researchers in Human-Computer Interaction (HCI) and Science and Technology Studies (STS) have explored how sociotechnical systems are shaped by, and in turn shape, institutional norms \cite{felt2017imaginaries,goni2025citizen,10.1093/0195180992.003.0007}.

In the context of platform governance, interface design has been shown to significantly influence user behavior and perceptions of fairness \cite{cai_content_2024}. Features such as the placement of buttons, the structure of forms, or the use of specific language can either encourage or deter users from engaging with regulatory mechanisms. The DSA explicitly recognizes this by requiring reporting mechanisms to be “easy to access” and “user-friendly” (Art. 16 DSA).
However, compliance is often assessed from a top-down legal perspective that overlooks the practicalities of design and technical implementation. 

Recent scholarship has emphasized the need for shared definitions and structured knowledge on dark patterns to better facilitate regulatory and academic dialogue. An ontology proposed by Gray et al. (2024) synthesizes existing taxonomies, offering a common language to address deceptive and coercive practices in digital design \cite{gray2024ontology}.
Studies like \emph{anonymized for review} have argued for evaluating of systems through regulation and usability-based design methodologies. In this paper, we contribute to this shift by examining how desigers can support the operationalization of legal compliance and digital humanism principles in digital services.

\section{Methodology}
We conducted a 1.5 hour workshop with nine \emph{anonymized for review} designers in December 2024. Participants were selected from various backgrounds, including UX design, interaction technology, computer science, and visual communication as described in Table \ref{tab:participants}. Their experience ranged from under one year to over ten years.

\begin{table}
\caption{Overview of Workshop Participants: Background, Experience, and Familiarity with the DSA}

\label{tab:participants}
\footnotesize
\renewcommand{\arraystretch}{1.15}
\begin{tabularx}{\textwidth}{l X l l l}
\toprule
\textbf{ID} & \textbf{Background} & \textbf{Experience} & \textbf{DSA Familiarity} & \textbf{Regulatory Context} \\
\midrule
P1 & UX Design               & 1--3 years   & Somewhat     & Yes \\
P2 & Informatics             & 3--5 years   & Not at all   & Yes \\
P3 & Media Production        & 5--10 years  & Somewhat     & Yes \\
P4 & Computer Science        & $<$1 year    & Not at all   & Yes \\
P5 & Visual Communication    & $>$10 years  & Somewhat     & No  \\
P6 & Interaction Technology  & 3--5 years   & Somewhat     & Yes \\
P7 & UX Design               & 5--10 years  & Not at all   & Yes \\
P8 & Visual Design           & $<$1 year    & Somewhat     & Yes \\
P9 & Design Management       & 1--3 years   & Not at all   & Yes \\
\bottomrule
\end{tabularx}
\end{table}

\begin{figure}
    \centering
    \includegraphics[width=1\linewidth]{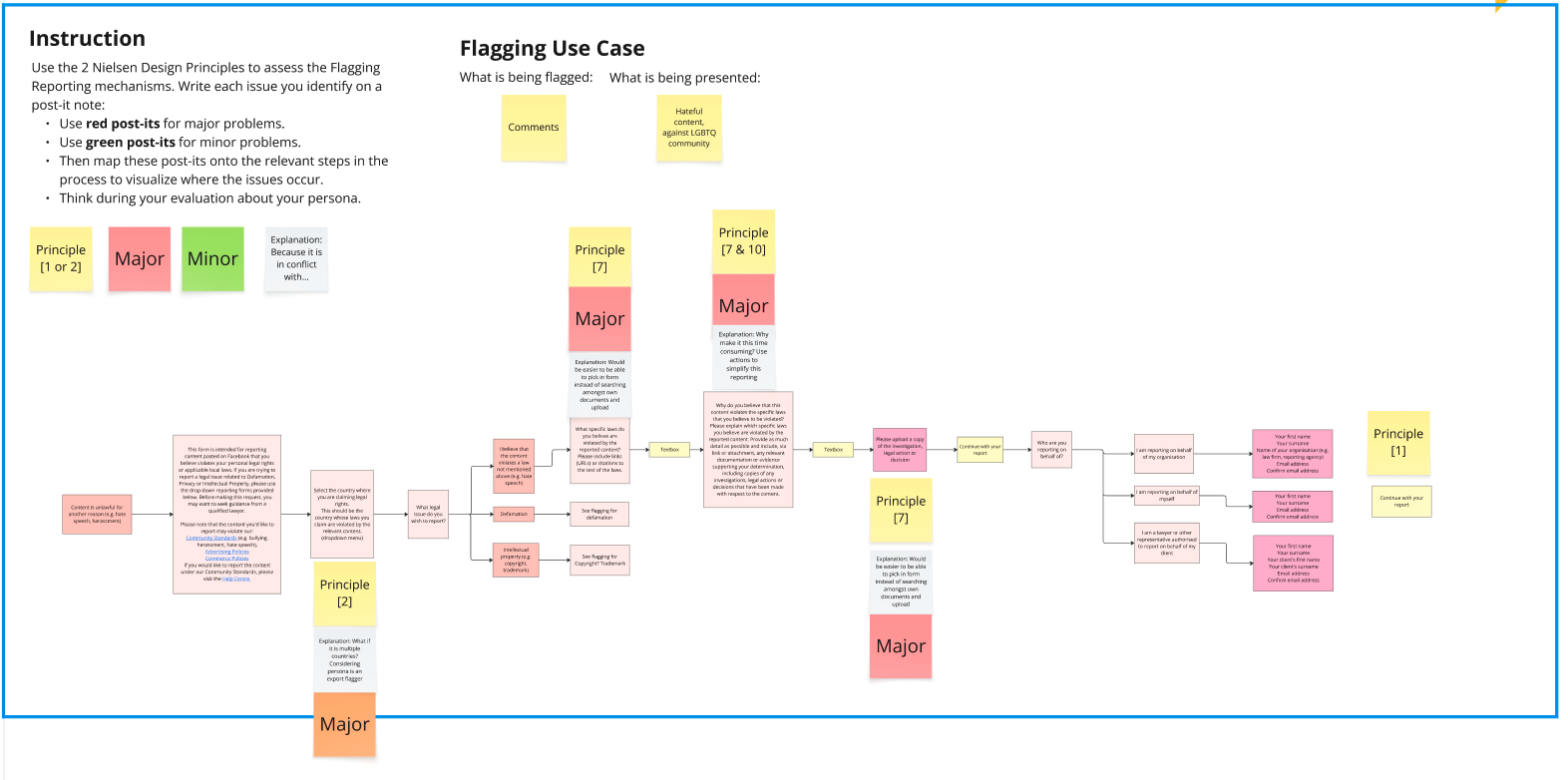}
    \caption{Participants evaluated the UI flows using Nielsen’s heuristics during the workshop through the lens of their persona (Bob = normal user/ Maria= expert user; Trusted Flagger).}
    \label{fig:enter-label}
\end{figure}

For an illustration of the wireframe evaluation process used in the workshopn to prompt discussion and design requirements, see Figure~\ref{fig:enter-label}. 

A survey via Menti indicated that five participants were only “somewhat familiar” with the DSA, while four had no familiarity. However, eight of the nine reported that their work intersected with at least one regulatory source of the Digital Decade (see column "Regulatory Context" in Table \ref{tab:participants} in the Appendix) such as GDPR, copyright law, or the AI Act.
The workshop was structured into three parts. First, participants were introduced to core concepts and workings of the DSA, important new roles like Trusted Flaggers (who are tasked with identifying and reporting illegal content according to Article 22 DSA) and its relevance for platform design. We focused in particular on Article 16, which sets out obligations for illegal content flagging mechanisms. We emphasized the dual requirement that such systems be both “easy to access” and “user-friendly.”

\begin{figure}
     \centering
     \includegraphics[width=1\linewidth]{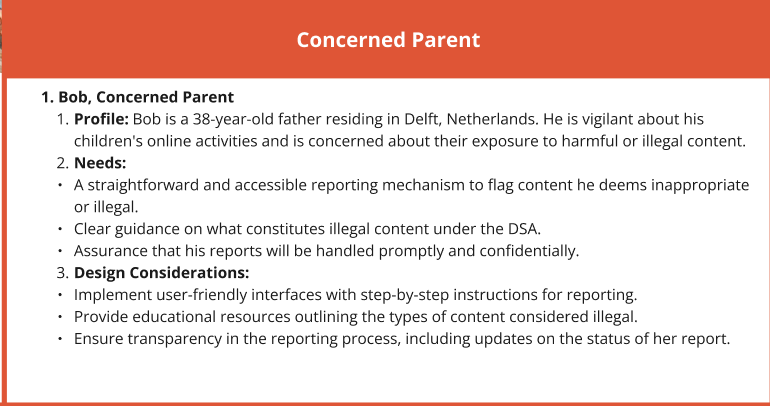}
\caption{Persona: Concerned Parent.\\
This persona represents a vigilant parent who is focused on protecting their children from harmful or illegal online content. The scenario emphasizes needs for straightforward, accessible reporting mechanisms, clear guidance on illegal content under the DSA, and transparency throughout the reporting process. Design considerations include step-by-step user instructions, educational resources, and prompt, confidential handling of reports.}

     \label{fig:enter-Marial}
 \end{figure}

\begin{figure}
    \centering
    \includegraphics[width=1\linewidth]{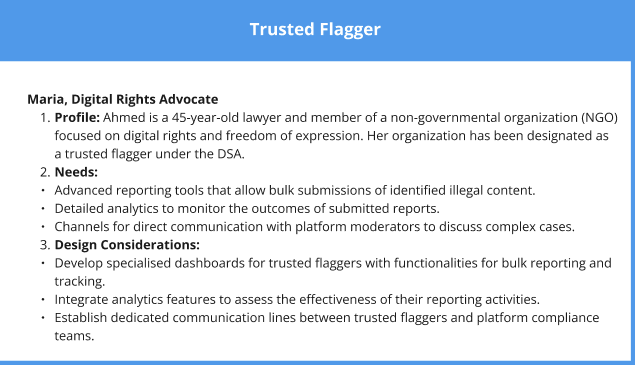}
 \caption{Persona: Trusted Flagger.\\
This persona illustrates an expert user, such as a digital rights advocate working for an NGO with Trusted Flagger status under the DSA. The scenario highlights needs for advanced reporting tools enabling bulk submissions, analytics to monitor outcomes, and direct communication channels with platform compliance teams. Design considerations include specialized dashboards, analytics integration, and dedicated support for handling complex cases.}

    \label{fig:enter-Bob}
\end{figure}

Second, participants engaged in a scenario-based walkthrough. We created two fictional reporting tasks: one involving the flagging of an illegal product offer on a marketplace, and another involving hate speech against the LGBTQ+ community. Each participant assumed the role of one of two personas — Bob, a concerned parent, or Maria, a Trusted Flagger as described in Figures \ref{fig:enter-Bob} and \ref{fig:enter-Marial}.

 Participants engaged with reporting interfaces through the lens of these personas. The first scenario featured Bob, a 38-year-old father from Delft, who is proactive about his children’s online safety and seeks a straightforward, accessible mechanism to flag content he deems harmful or illegal. His needs centered on clear guidance regarding what constitutes illegal content under the DSA, as well as assurance of confidentiality and transparency throughout the reporting process. 
 The second scenario introduced Maria, a digital rights advocate and lawyer whose NGO has been designated as a Trusted Flagger under the DSA. In this expert role, her requirements included access to advanced reporting tools for bulk submission, analytics to assess outcomes, and dedicated channels for direct communication with platform compliance teams. By evaluating the same interface from these distinct perspectives, participants were able to surface both general and specialized challenges in the design of DSA-compliant reporting mechanisms, reflecting the diversity of end-user experiences and regulatory expectations.
We provided wireframes based on \emph{redacted for review} modeled after Facebook’s reporting interface to maintain ecological validity.

Third, participants evaluated the UI flows using Nielsen’s heuristics \cite{nielsen_heuristic_1990,nielsen_enhancing_1994,nielsen_finding_1992}
 and categorized issues as major or minor using color-coded post-its (major=red, minor=green). They were also encouraged to reflect on their decisions (explanation=white) how the system aligned with Nielsen's design principles (design principle=yellow). Qualitative insights were collected via Miro boards and a follow-up discussion.
To guide participants in their evaluation of platform interfaces, we adopted Nielsen’s 10 usability heuristics for user interface design \cite{nielsen_heuristic_1990,nielsen_enhancing_1994,nielsen_finding_1992}
. These heuristics are a widely accepted set of principles used in Human-Computer Interaction (HCI) to identify usability problems and assess design quality. They provide a structured yet flexible framework for evaluating whether an interface aligns with good design practices. Each heuristic targets a core dimension of user experience, from feedback and control to error prevention and aesthetic simplicity, for example
\emph{Visibility of System Status} under which The system should always keep users informed about what is happening through timely and relevant feedback or \emph{Match Between System and the Real World} where the interface should speak the user’s language, using familiar concepts, metaphors, and conventions.
These heuristics allowed workshop participants to identify friction points in current flagging interfaces, categorize them by severity, and reflect on the regulatory implications of interface failures. Lastly, we connect the findings from the workshop with principles from the digital humanism manifesto to connect core familiarities.

\section{Findings}
Participants identified a series of structural and interactional problems with the existing flagging workflows. One of the most common observations concerned discoverability. Several designers noted that the reporting option was “hidden” or “buried” within multiple layers of the interface. This problem was especially acute on mobile, where button placement and visual hierarchy reduced the salience of the feature as highlighted in the second flagging example capturing the challenge of hate speech against the LGBTQ+ community in user comments as highlighted in \emph{redacted for review}.

Participants also criticized the tone of the interface, describing it as impersonal and using too much legal jargon. In scenarios involving sensitive content such as hate speech, designers emphasized the need for emotionally responsive language and supportive cues. As one participant noted, “the system feels like it’s talking to a lawyer, not a person who’s just been harmed.”
A third major theme was legal ambiguity. Options such as “intellectual property violation” or “hate crime” were not clearly explained, and participants reported uncertainty about how to choose the right category. This was seen as a barrier for both scenarios, especially when platforms required citation of specific legal statutes in the complaint process.

Lack of feedback was another key issue. Designers were highlighting the absence of confirmation messages or progress indicators. Some suggested that this opacity could discourage user engagement. Finally, the uniformity of the reporting process was flagged as a limitation. All users 
regardless of their expertise 
were forced to follow the same linear path, despite having different needs and levels of familiarity.

\section{Discussion}
Our findings resonate with previous research on the design of reporting mechanism and dark patterns \cite{wagner_regulating_2020,gray_legal_2024,di2020ui}
\emph{redacted for review}. We argue that current reporting mechanisms often fall short of the DSA’s mandate for user-centered design. Rather than supporting empowerment and agency, they create friction and confusion – also for designers. This is problematic because such design gaps can lead to inconsistent or vague implementation of the DSA’s requirements, ultimately making the reporting process more confusing for users and undermining the regulation’s objectives of transparency, accessibility, and user protection.

Our findings show that participatory legal UX methods encourage critical discussions between legal and design professionals, making them essential for advancing the main aims of Digital Humanism. 
By engaging designers directly with legal frameworks and regulatory goals, participatory workshops transform abstract legal mandates into design experiences. This practice surfaces not only technical and usability challenges, but also underlying value tensions such as the balance between legal certainty and user empowerment, or between standardized processes and inclusivity.

Participatory prototyping can serve as a speculative enactment practice, allowing designers to visualize regulatory impacts and privacy considerations early in the design process \cite{nelissen2022rationalizing}.

The workshop revealed that a single, standardized reporting flow made the system consistent, but excluded users with less legal knowledge or different motivations (e.g., a parent vs. a Trusted Flagger). Some participants expressed that the design did not account for varying expertise or the emotional state of users, limiting inclusivity and accessibility.

In line with the Vienna Manifesto on Digital Humanism, our work acknowledges that while designers, legal scholars, and other professionals play a crucial role in shaping technology, these technologies in turn influence and reshape our behaviors, norms, and expectations. This reciprocal relationship underscores the importance of not only theorizing about regulation, but also testing and refining it through design experiments and practical evaluation. Such iterative approaches are essential to understanding how regulatory goals and human values are realized or challenged in real-world contexts.

The process of co-designing DSA-compliant interfaces provided a forum for interdisciplinary sense-making, where the legal requirements are brought to the attention of designers and explained through design examples. Designers did not act as passive implementers, but as co-creators who interpreted, questioned, and reimagined what “user-friendliness,” or“accessibility” should mean in practice.

Participatory legal UX thus embodies several Digital Humanism principles:

\textbf{Democracy and inclusion}: Workshops create space for diverse perspectives, breaking down disciplinary silos and exploring digital experiences of experts and everyday users.

\textbf{Transparency and accountability}: By involving practitioners in regulatory translation, the process makes legal requirements legible and actionable, supporting both fairness and user agency.

\textbf{Interdisciplinarity}: The collaborative nature of expert workshops with a specific group of participants called for by the Manifesto, bringing together law and design.

\textbf{Effective regulations, rules and laws, based on public discourse}: Our workshop emphasizes the need to also provide solutions through effective and compliant design by leveraging an interdisciplinary audience.

Our study supports the growing consensus that designers should play a central role not only in implementing but also in shaping digital policy solutions and legal details. As highlighted by Design for Values approaches \cite{vandenhoven2015,vandePoel2024}, design activities surface value tensions, such as empowerment versus standardization, which are fundamental to regulatory debates. Participatory prototyping and design \cite{chock2020design} and testing enable a translation of high-level regulatory aims into actionable, context-sensitive requirements. This is particularly important as platform regulation increasingly intersects with AI systems (see also Article 14 AI Act), where poorly designed human-AI collaborations can perpetuate or exacerbate bias, rather than mitigate it \cite{constantino2024digitaldecade}. 

We therefore propose that involving designers in policy and regulatory related prototyping, as well as the ongoing testing of regulatory effects, represents a promising direction for both research, legal design and regulatory practice.
Finally, our findings suggest that participatory legal UX can act as a catalyst for broader societal engagement with digital regulation. By embedding processes of collective sense-making into the development of compliant digital systems, we move towards a digital society that not only is aware of legal obligations, but upholds and advances human rights also in digital spaces. As such, participatory legal UX is not merely a technical tool, but a strategy for enacting Digital Humanism in platform governance.

\section{Conclusion}
Designers are uniquely positioned to operationalize regulation in the digital space. As this case study demonstrates, participatory design methods can surface hidden tensions in the implementation of complex laws like the DSA and generate actionable strategies for improvement. More importantly, they can empower designers to take an active role in shaping regulatory compliance, rather than being passive recipients of legal obligations.

As new regulations emerge globally, the ongoing involvement of designers, as both implementers and co-creators of law and codified legal realities, will be vital to ensure that digital systems reflect not just the letter, but also the spirit and values, of society. Future work should explore methods for deeper integration of design expertise into regulatory development, prototyping, and evaluation.

We conclude by calling for greater interdisciplinary collaboration between legal scholars, policymakers, engineers, and design practitioners. As the digital regulation landscape evolves, so too must our methods for enacting it. Legal UX offers a promising path forward one that recognizes that how we experience the law is just as important as what the law says.

\newpage

\end{document}